\documentclass[10pt, onecolumn]{article}
\usepackage[breaklinks=true]{hyperref}
\usepackage{amsmath}
\usepackage{authblk}
\usepackage{float}
\usepackage[margin=1in]{geometry}
\usepackage{graphicx}
\usepackage{tikz}
\usepackage[utf8]{inputenc}
\usepackage[numbers]{natbib}
\usepackage{tabularx}
\usepackage{url}
\newcolumntype{n}{X}
\newcolumntype{s}{>{\hsize=.35\hsize}X}

\restylefloat{table}

\title{Google COVID-19 Vaccination Search Insights:\\Anonymization Process Description (version 1.0)}

\author{Shailesh Bavadekar}
\author{Adam Boulanger}
\author{John Davis}
\author{Damien Desfontaines}
\author{Evgeniy Gabrilovich}
\author{Krishna Gadepalli}
\author{Badih Ghazi}
\author{Tague Griffith}
\author{Jai  Gupta}
\author{Chaitanya Kamath}
\author{Dennis Kraft}
\author{Ravi Kumar}
\author{Akim Kumok}
\author{Yael Mayer}
\author{Pasin Manurangsi}
\author{Arti Patankar}
\author{Irippuge Milinda Perera}
\author{Chris Scott}
\author{Tomer Shekel}
\author{Benjamin Miller}
\author{Karen Smith}
\author{Charlotte Stanton}
\author{Mimi Sun}
\author{Mark Young}
\author{Gregory Wellenius}

\affil{covid-19-search-trends-feedback@google.com}

\date{July 1, 2021}

\begin{document}
\maketitle
\begin{abstract}
This report describes the aggregation and anonymization process applied to the COVID-19 Vaccination Search Insights~\cite{vaccination}, a publicly available dataset showing aggregated and anonymized trends in Google searches related to COVID-19 vaccination. The applied anonymization techniques protect every user’s daily search activity related to COVID-19 vaccinations with $(\varepsilon, \delta)$-differential privacy for $\varepsilon = 2.19$ and $\delta = 10^{-5}$.
\end{abstract}

\section{Introduction}

The COVID-19 Vaccination Search Insights is a publicly available dashboard~\cite{vaccination} and dataset~\cite{vaccinationdataset} that shows aggregated, anonymized trends in Google searches related to COVID-19 Vaccination. The data provides a weekly time series for each region showing the relative volume of searches across 3 categories:
\begin{itemize}
    \item \textbf{COVID-19 Vaccination:} All searches related to COVID-19 vaccination, indicating overall search interest in the topic.
    \item \textbf{Vaccination Intent:} Searches related to eligibility, availability, and accessibility of vaccines.
    \item \textbf{Safety and Side Effects:} Searches related to the safety and side effects of the vaccines.
\end{itemize}

We are making this data available because we have heard from public health organizations and researchers that they want access to localized and timely data about what information their communities are seeking, so they can better understand the vaccination needs of their communities and tailor their communication to people not yet vaccinated. 

To ensure strict privacy standards, all published data is aggregated and anonymized, protecting each user’s search activity on a given day using a differentially private mechanism. No personal data is included in the dataset.

The Vaccination Search Insights data reflect the (normalized) volume of Google searches that may be associated with a particular COVID-19 vaccination search topic. For this purpose, we count the weekly number of searches relating to that topic within a given geographic region, and normalize this number based on the total search activity in that region. The resulting data is a weekly series showing the relative frequency of searches for that particular topic in a particular region. The data covers the recent period and we plan to gradually expand its range as part of regular updates.
 
Similar to the Google COVID-19 Community Mobility Reports~\cite{mobilitytechreport} and Symptom Search Dataset~\cite{symptomtechreport}, and as explained in greater technical detail below, the anonymization process of the Vaccination Search Insights data is based on differential privacy~\cite{dp}, which is a well established approach for producing data that provides formal privacy guarantees. For this purpose, we intentionally perturb our data by adding random noise, and then drop data points that are deemed to be unreliable. The Vaccination Search Insights are designed to maintain the privacy of our users while releasing aggregated and anonymized data that is as accurate and useful as possible. 

The remainder of this report is structured as follows. First, we introduce some basic concepts and terminology of the COVID-19 Vaccination Search Insights. We then explain how differential privacy is used to produce anonymized aggregates. Finally, we elaborate on how the published data is built from the anonymized aggregates.

\section{Definitions}

The following definitions explain some common terms and concepts that are used throughout this report.

\begin{itemize}
    \item \textbf{User:} A user who issued a search query on Google Search.
    \item  \textbf{Search Categories:} We classify Google search queries according to the following 4 categories:
    \begin{itemize}
        \item \textbf{[A0] Any}: Contains all search queries.
        \item \textbf{[A1] Vaccination Intent:} Contains searches related to eligibility, availability, and accessibility of COVID-19 vaccines.
        \item \textbf{[A2] Safety and Side Effects:} Contains searches related to the safety and side effects of COVID-19 vaccines.
        \item \textbf{[A3] Other:} Contains all searches related to COVID-19 vaccination that are not classified as Vaccination Intent or Safety and Side Effects.\footnote{The COVID-19 Vaccination category, which indicates the overall interest in the topic, is derived from the union of the Vaccination Intent, Safety and Side Effects and Other categories as described in the Post-processing section.}
    \end{itemize}
    The classification is done via a machine learning model that was trained in a semi-supervised fashion.
    \item \textbf{Geographic Regions:} Search queries are aggregated based on the geographic region where they are issued. We provide data for four geographic granularity levels: \textbf{Country}\footnote{Country level data is not generated directly, instead we aggregate it from anonymized State level data. Thus, we avoid spending privacy budget at the Country level.} and three levels of subdivision within each country. In the US, these three levels are \textbf{State}, \textbf{County}, and \textbf{Postal Code}, which we use as our running example in the remainder of this paper. All reported geographic regions have an area of at least $3\textrm{km}^2$.
    \item \textbf{County and Postal Code Types:} We distinguish between three types of counties based on the size of their population, relying on the 2018 US Census Bureau data:
    \begin{itemize}
        \item \textbf{Small:} Counties with a population of fewer than 100,000.
        \item \textbf{Medium:} Counties with a population between 100,000 and 500,000.
        \item \textbf{Large:} Counties with a population of greater than 500,000.
    \end{itemize}
    Moreover, we label postal codes with the same type, i.e., Small, Medium, and Large, as the county they are part of.\footnote{If a postal code region stretches across multiple counties, we match its type with the county with which it shares most of its addresses.} Note that this classification only depends on the population size of the county, not on the population size of the postal code.

    \item \textbf{Differential Privacy:} Let $A$ be a randomized algorithm that takes a dataset as input. We say that $A$ is $(\varepsilon, \delta)$-differentially private if for any pair of neighboring datasets $D_1$ and $D_2$ and for all sets $S$ of the possible outputs of $A$, it holds that:
    \[\textrm{Pr}[A(D_1) \in S] \leq \textrm{exp}(\varepsilon) \cdot \textrm{Pr}[A(D_2) \in S] + \delta.\]
    In the context of this report, we consider $D_1$ and $D_2$ neighboring datasets if $D_1$ can be obtained from $D_2$ by adding or removing a single user’s search activity on a given day.
\end{itemize}

\section{Anonymization Process}

The COVID-19 Vaccination Search Insights are based on anonymized search query counts. The counts are grouped by the week in which a query was made, the region from which it was issued and its category. This means that each count can be uniquely associated with a $<$week, region, category$>$ triple.

As part of the anonymization process, we limit each user’s daily contribution to the search query counts and add appropriately scaled Gaussian noise. The resulting metrics protect each user’s daily search activity with $(\varepsilon, \delta)$-differential privacy, where $\varepsilon = 2.19$ and $\delta = 10^{-5}$.

\subsection{Contribution Bounding and Constraints}

A user’s contributions to the vaccination search counts are subject to the following constraints:

\begin{itemize}
    \item \textbf{[X1] Cross-Count Bounding:} A user can contribute to at most 1 count per day, geographic level, and category.
    \item \textbf{[X2] Per-Count Bounding:} A user can increment each count by at most 1 per day.
    \item \textbf{[X3] Small Postal Code Constraint:} We do not report on Postal Code counts of type Small.\footnote{The motivation behind this constraint is to reduce the relative error differential privacy introduces to County level counts. Using population size as a proxy for the relative error, the omitted counts free up privacy budget at Postal Code level that we then use to reduce the noise added at the County level.}
    \item \textbf{[X4] County / Postal Code Synchronization:} A user cannot contribute to counts of different County and Postal Code types on the same day.\footnote{This constraint is intended to keep our privacy budget accounting consistent, which is necessary because we use different noise parameterizations for different County and Postal Code types.}
\end{itemize}

Similar to the process described by Wilson et al.~\cite{privatesql}, we arbitrarily drop search queries that violate these constraints. Note that none of these contribution bounds apply to more than a single day, so we can consider each day’s data separately when enforcing them.

To give an example, imagine a user who makes the following searches during the 10\textsuperscript{th} week of 2021:

\begin{itemize}
    \item \textbf{Search 1:} Unrelated to COVID-19 Vaccinations, made on March 9\textsuperscript{th}, 2021 from Postal Code 94103, San Francisco County, California.
    \item \textbf{Search 2:} Related to the safety of COVID-19 Vaccinations, made on March 9\textsuperscript{th}, 2021 from Postal Code 95023, San Benito County, California.
    \item \textbf{Search 3:} Related to the eligibility of COVID-19 Vaccinations, made on March 11\textsuperscript{th}, 2021 from Postal Code 95023, San Benito County, California.
\end{itemize}

Table~\ref{tab:prebounding} shows how the user’s searches are mapped to the counts. Note that the resulting mapping violates some of the constraints we established earlier. To resolve these conflicts, we go step by step through the constraints and drop contributions. The choice of contributions we drop and the order in which we drop them is arbitrary and only intended for illustration purposes. We begin with March 9\textsuperscript{th}, when the user made search queries 1 and 2:

\begin{table}[H]
    \centering
    \begin{tabular}{| l | l | l | l |}
        \hline
        Count ($<$week, category, region$>$)            & Search                            & Geographic Level  & Region Type\\
        \hline
        $<$10, Any, 94103$>$                            & 1$^\textrm{X4}$                   & Postal Code       & Large\\
        $<$10, Any, 95023$>$                            & 2$^\textrm{X1}$, 3$^\textrm{X3}$  & Postal Code       & Small\\
        $<$10, Any, San Francisco$>$                    & 1$^\textrm{X4}$                   & County            & Large\\
        $<$10, Any, San Benito$>$                       & 2$^\textrm{X1}$, 3                & County            & Small\\
        $<$10, Any, California$>$                       & 1, 2$^\textrm{X2}$, 3             & State             & N/A\\
        $<$10, Safety and Side Effects, 95023$>$        & 2$^\textrm{X3}$                   & Postal Code       & Small\\
        $<$10, Safety and Side Effects, San Benito$>$   & 2                                 & County            & Small\\
        $<$10, Safety and Side Effects, California$>$   & 2                                 & State             & N/A\\
        $<$10, Vaccination Intent, 95023$>$             & 3$^\textrm{X3}$                   & Postal Code       & Small\\
        $<$10, Vaccination Intent, San Benito$>$        & 3                                 & County            & Small\\
        $<$10, Vaccination Intent, California$>$        & 3                                 & State             & N/A\\
        \hline
  \end{tabular}
  \caption{Mapping of search queries 1, 2, and 3 to the respective counts before constraint violations are resolved. Searches that will be dropped because of a violation are marked with X1, X2, X3, or X4 according to the violated constraint. Columns 3 and 4 provide auxiliary information for identifying violations.}
  \label{tab:prebounding}
\end{table}

\begin{itemize}
    \item {[X1] Cross-Count Bounding:}
    \begin{itemize}
        \item Conflict: In the Any category at the Postal Code level the user contributed from both 94103 and 95023. In the Any category at the County level, the user contributed from both San Francisco and San Benito.
        \item Fix: Drop search 2 from $<$Week 10, Any, 95023$>$ and $<$Week 10, Any, San Benito$>$.
    \end{itemize}
    \item {[X2] Per-Count Bounding:}
    \begin{itemize}
        \item Conflict: In the Any category, at State level, the user contributed two searches.
        \item Fix: Drop search 2 from $<$Week 10, Any, California$>$.
    \end{itemize}
    \item {[X3] Small Postal Code Constraint:}
    \begin{itemize}
        \item Conflict: The user contributed to Postal Code 95023, which is Small.
        \item Fix: Drop search 2 from $<$Week 10, Safety and Side Effects, 95023$>$.
    \end{itemize}
    \item {[X4] County / Postal Code Synchronization:}
    \begin{itemize}
        \item Conflict: The user contributed to San Francisco and Postal Code 94103, which are both Large, in the Any category. The user also contributed to San Benito, which is Small, in the Safety and Side Effects category.
        \item Fix: Drop search 1 from $<$10, Any, 94103$>$ and $<$10, Any, San Francisco$>$.\footnote{We already resolved a conflict between 94103 and 95023 in the Any category in the cross-count bounding step by removing the contributions to 95023, but that still leaves a synchronization conflict between the remaining Any values for 94103 and the Safety and Side Effects values for 95023. In light of both conflicts, it might have been more efficient to solve both by dropping the 94103 values. Our goal in this example is to demonstrate all possible types of constraint violations. It does not directly correspond to the way we resolve conflicts in practice.}
     \end{itemize}
\end{itemize}

Next, we consider March 11\textsuperscript{th} when the user made search query 3. Because this query was made on a different day, it does not conflict with 1 and 2. Moreover, a single query in a day can only violate the Small Postal Code Constraint. As it turns out, this constraint is indeed violated and therefore we drop search 3 from $<$Week 10, Any, 95023$>$ and $<$Week 10, Vaccination Intent, 95023$>$. Table~\ref{tab:postbounding} shows the resulting contributions, which now satisfy all constraints:

\begin{table}[H]
    \centering
    \begin{tabular}{| l | l | l | l |}
        \hline
        Count ($<$week, category, region$>$)            & Search    & Geographic Level  & Region Type\\
        \hline
        $<$10, Any, San Benito$>$                       & 3         & County            & Small\\
        $<$10, Any, California$>$                       & 1, 3      & State             & N/A\\
        $<$10, Safety and Side Effects, San Benito$>$   & 2         & County            & Small\\
        $<$10, Safety and Side Effects, California$>$   & 2         & State             & N/A\\
        $<$10, Vaccination Intent, San Benito$>$        & 3         & County            & Small\\
        $<$10, Vaccination Intent, California$>$        & 3         & State             & N/A\\
        \hline
  \end{tabular}
  \caption{Mapping of search queries 1, 2, and 3 to the respective counts after constraint violations are resolved. Columns 3 and 4 provide auxiliary information for identifying violations.}
  \label{tab:postbounding}
\end{table}

\subsection{Noise Addition and Privacy Budget Accounting}

The standard deviation $\sigma$ of the Gaussian noise we add to each search query count depends on the population size type and geographic level of the region as well as the search category. Table~\ref{tab:stdv} lists all standard deviation values we use:

\begin{table}[H]
    \centering
    \begin{tabular}{| l | l | l | l |}
        \hline
        Population  & Level         & Search Category   & Standard Deviation\\
        \hline
        Large       & Postal Code   & A0                & $\sigma = 35.0$\\
        Large       & Postal Code   & A1, A2, A3        & $\sigma = 3.25$\\
        Large       & County        & A0                & $\sigma = 180.0$\\
        Large       & County        & A1, A2, A3        & $\sigma = 20.0$\\
        Medium      & Postal Code   & A0                & $\sigma = 40.0$\\
        Medium      & Postal Code   & A1, A2, A3        & $\sigma = 3.5$\\
        Medium      & County        & A0                & $\sigma = 100.0$\\
        Medium      & County        & A1, A2, A3        & $\sigma = 8.0$\\
        Small       & County        & A0                & $\sigma = 28.0$\\
        Small       & County        & A1, A2, A3        & $\sigma = 3.21$\\
        N/A         & State         & A0                & $\sigma = 450.0$\\
        N/A         & State         & A1, A2, A3        & $\sigma = 35.0$\\
        \hline
  \end{tabular}
  \caption{Standard deviation of the Gaussian noise added to the search query counts.}
  \label{tab:stdv}
\end{table}

To establish the total $\varepsilon$ and $\delta$ provided by the Gaussian noise, we consider 3 cases based on the difference between the neighbouring data sets $D_1$ and $D_2$ from the definition of $(\varepsilon, \delta)$-differentially privacy (i.e. the contributions that are in $D_1$ but not $D_2$ or vice versa):

\begin{itemize}
    \item \textbf{Case 1:} The differing contributions at County and Postal Code level are from Large regions.
    \item \textbf{Case 2:} The differing contributions at County and Postal Code level are from Medium regions.
    \item \textbf{Case 3:} The differing contributions at County level are from Small counties and there are no differing contributions at Postal Code level.
\end{itemize}

As a result of the County / Postal Code Synchronization and the Small Postal Code Constraint, exactly one of the three cases applies. For each case, we use our open source privacy accounting library~\cite{dpacclib}, to compute $\varepsilon$ and $\delta$ values that satisfy
\[\textrm{Pr}[A(D_1) \in S] \leq \textrm{exp}(\varepsilon) \cdot \textrm{Pr}[A(D_2) \in S] + \delta.\]

\begin{itemize}
    \item Case 1: The Cross-Count Bounding and Per-Count Bounding implies that the distinguishing contributions are made to at most 12 Gaussian Mechanisms with standard deviations
    \[\delta_{1,...,12} = (3.25, 3.25, 3.25, 35.0, 20.0, 20.0, 20.0, 180.0, 35.0, 35.0, 35.0, 450.0)\]
    and a sensitivity of 1 each. For a fixed  $\delta = 10^{-5}$, this yields $\epsilon \approx 2.186$.
    \item Case 2: Similar to the previous case, the distinguishing contributions are made to at most 12 Gaussian Mechanisms with standard deviations
    \[\delta_{1,...,12} = (3.5, 3.5, 3.5, 40.0, 8.0, 8.0, 8.0, 100.0, 35.0, 35.0, 35.0, 450.0)\]
    and sensitivities 1 each. For a fixed $\delta = 10^{-5}$, this yields $\epsilon \approx 2.187$.
    \item Case 3: The distinguishing contributions are now made to at most 8 Gaussian Mechanisms (because Postal Code level contributions are omitted) with standard deviations
    \[\delta_{1,...,12} = (3.21, 3.21, 3.21, 28.0, 35.0, 35.0, 35.0, 450.0)\]
    and sensitivities 1 each. For a fixed  $\delta = 10^{-5}$, this yields $\epsilon \approx 2.186$.
\end{itemize}

Independent of which case applies, we can bound the overall $\varepsilon$ by
\[\varepsilon \leq \textrm{max}\{\varepsilon_1, \varepsilon_2, \varepsilon_3\} \leq 2.19,\]
which yields the advertised $(2.19, 10^{-5})$-differential privacy guarantee.

\section{Post-processing}

We report search trends in the following categories:
\begin{itemize}
    \item \textbf{[C1] Vaccination Intent:} A subset of queries about COVID-19 vaccination that are related to eligibility, availability, and accessibility of vaccines.
    \item \textbf{[C2] Safety and Side Effects:} A subset of queries about COVID-19 vaccination that are related to the safety and side effects of vaccines.
    \item \textbf{[C3] COVID-19 Vaccination:} All queries related to COVID-19 vaccination.
\end{itemize}

Obviously, these 3 categories are not mutually exclusive. To make it easier to implement  contribution bounding, we chose to manipulate categories that are mutually exclusive, which led us to introduce categories A1, A2, and A3 earlier in this document. We derive the final categories C1, C2, and C3 from those introduced earlier like this:
\begin{itemize}
    \item Count(C1) = Count(A1)
    \item Count(C2) = Count(A2)
    \item Count(C3) = Count(A1) + Count(A2) + Count(A3)
\end{itemize}

\subsection{Normalizing the Counts}

The statistics described in the previous step are in the form of raw counts with added differential privacy (Gaussian) noise. The first transformation we apply is to normalize these counts by the total search volume in each region, i.e., Count(A0), which will offer a probabilistic interpretation. For example, Count(C3) / Count(A0) can serve as a proxy for the probability of a random query to be related to COVID-19 vaccination. We refer to the resulting ratios as \textbf{normalized popularity} of each category. This normalization is performed at every level of geographic granularity (Postal Code, County, etc.)
 
Since the normalized popularity values have a probabilistic interpretation, they can be compared across time intervals and geographic regions.

\subsection{Removing Unreliable Data}

In some geographic regions, the noise added to the data by the differentially private mechanism can introduce a disproportionate amount of uncertainty. Typically, this happens when the respective count is 0 or small. To limit the amount of uncertainty, we only keep the noisy value of a metric if it has a chance of 80\% or more to be within 15\% of its original value, i.e., the value before the addition of noise.

More precisely, let $X$ be a noisy count obtained after adding Gaussian noise to the original count $x^*$. Note that $x*$ is not noisy but still subject to contribution bounding. Similarly, let $Y$ be the corresponding noisy normalization count computed from the raw count $y*$. To decide whether the data associated with $X$ will be kept or dropped:

We first compute a confidence interval $[l, r]$ based on $X$ and $Y$ that contains $x^* / y^*$ with a probability of at least 80\%. We then keep the data if the following inequalities all hold:
\begin{itemize}
    \item $X / Y > 0$
    \item $(X / Y)-l \leq 0.15 \cdot X / Y$
    \item $r - (X / Y) \leq 0.15 \cdot X / Y$
\end{itemize}
Otherwise, we drop this data point.

The confidence interval $[l, r]$ is entirely based on the anonymized counts $X$ and $Y$. As a result, no privacy budget is spent on removing unreliable data.

\subsection{Empty Data Points}

Note that the volume of search traffic related to the different topical categories has varied widely throughout the course of the pandemic. Consequently, on some weeks some small locations might not have received any search traffic in a given category. In other cases, the volume of search traffic was small, and the corresponding data point was removed as unreliable (cf. the previous section). As a result, some weekly data points may be missing, especially in less populated regions, and some regions might exhibit data sparseness. Although data points that have not been filtered out are generally reliable, we remove regions that have extreme data sparsity, namely, those that have 3 data points or fewer during the period January - May 2021 (which corresponds to over 85\% of data points missing).

\subsection{Scaling the Normalized Counts}

The normalized counts can be seen as a proxy for the probability of a query to belong to a given category, and these numbers are generally very small (users search on a large variety of topics, and a relatively small fraction of the queries are related to COVID-19 vaccination, let alone its sub-categories). To make the numbers easier to manipulate and visualize, we scale the normalized popularity values as explained below.

First, we identify the maximum value of the normalized popularity of the COVID-19 vaccination category (C3), across the entire initially-published US national time series. We scale this maximum value to 100. You can think of this as the highest weekly interest in COVID-19 vaccination, at the US national level, at the time of the initial data release.

Using the same scaling factor, we linearly scale all the other values, across regions, categories, and time (the resulting values may be higher or lower than 100). We store the scaling factor computed in the initial release and use it to scale the values across all regions and categories in subsequent releases.

Observe that the scaling factor is determined purely based on the noisy search counts and normalization counts, so no privacy budget is spent on its computation. Observe also that a single scaling factor is applied across data from all locations, weeks, and categories. As a result, the scaled data points are still comparable across all of these dimensions.
\bibliographystyle{unsrt}
\bibliography{references}
\end{document}